\documentclass[12pt,preprint]{aastex}
\def\tO{{\tilde{\Omega}}}
\def\bv{{\bf v}}
\def\bB{{\bf B}}
\def\bnabla{{\bf \nabla}}
\def\<{\langle}
\def\>{\rangle}
\def\alf{Alfv\'en }

\def\va{{\bf v}_{\rm A}}
\def\bk{{\bf k}}

\def\eps{\epsilon}
\def\sA{{\mathcal{A}}}
\def\sB{{\mathcal{B}}}
\def\sC{{\mathcal{C}}}
\def\sD{{\mathcal{D}}}
\def\sF{{\mathcal{F}}}
\def\sG{{\mathcal{G}}}
\def\order{{\mathcal{O}}}

\shortauthors{}
\shorttitle{Kerr MRI}
\begin{document}
\title{The Magnetorotational Instability in the Kerr Metric}

\author{Charles F. Gammie\altaffilmark{1,2}}

\affil{Center for Theoretical Astrophysics, University of 
Illinois, Urbana, IL 61801 USA}

\email{gammie@uiuc.edu}

\altaffiltext{1}{Department of Astronomy}
\altaffiltext{2}{Department of Physics}

\begin{abstract}

The magnetorotational instability (MRI) is the leading candidate for
driving turbulence, angular momentum transport, and accretion in
astrophysical disks.  I consider the linear theory of the MRI in a thin,
equatorial disk in the Kerr metric.  I begin by analyzing a mechanical
model for the MRI that consists of two point masses on nearly circular
orbits connected by a spring.  I then develop a local Cartesian
coordinate system for thin, equatorial Kerr disks.  In this local model
general relativistic effects manifest themselves solely through changes
in the Coriolis parameter and in the tidal expansion of the effective
potential.  The MRI can be analyzed in the context of the local model
using nonrelativistic magnetohydrodynamics, and the growth rates agree
with those found in the mechanical model.  The maximum growth rate
measured by a circular orbit observer differs from a naive estimate
using Newtonian gravity by a factor that varies between $1$ and $4/3$
for all radii and for all $a/M$.

\end{abstract}


\section{Introduction}

Balbus \& Hawley (1991; hereafter BH) discovered a local instability of
weakly magnetized accretion disks, studied earlier in the context of
magnetized Couette flow by \cite{vel59} and others.  This instability is
widely known as the magnetorotational instability (MRI).  The
instability conditions (see \citealt{bh98}) are easily satisfied, so
almost every ionized astrophysical disk is likely subject to the
instability.  Numerical experiments have demonstrated that the MRI can
initiate and sustain magnetohydrodynamic (MHD) turbulence in disks (e.g.
\citealt{hgb95}).  They have also shown that MHD turbulence in disks
transports angular momentum outward.  MRI initiated turbulence is
therefore a plausible (arguably the most plausible) candidate for
driving accretion in astrophysical disks.

Disks around black holes are also likely subject to the MRI.  The
development of the MRI in these disks is of fundamental interest.  But
it is also of practical interest because numerical models of
relativistic disks are now being developed \citep{dvh03, gmt03, dvhk03,
gsm03}.  In particular, one would like to know if the maximum growth
rate is very different from the $(3/4) (G M/r^3)^{1/2}$ expected from
analyses of Newtonian disks.

In this short paper I will study the MRI in thin, equatorial disks in
the Kerr metric.  I will use units such that $GM = c = 1$, where $M$ is
the black hole mass, and Boyer-Lindquist coordinates $t,r,\theta,\phi$,
where
\begin{equation}
ds^2 = - \left( 1 - \frac{2\,r}{\Sigma}\right)\,dt^2 +
  \frac{\Sigma}{\Delta}\,dr^2 +
    \Sigma\,d\theta^2 + \frac{A\,{\sin^2
      {\theta}}}{\Sigma}\,d\phi^2 - \frac{4\,a\,r\,{\sin^2
        {\theta}}}{\Sigma}\,d\phi\,dt,
\end{equation}
and $\Sigma\equiv r^2+a^2\cos^2{\theta}$, $\Delta\equiv r^2 - 2 r +
a^2$ and $A\equiv (r^2+a^2)^2- a^2\Delta\sin^2\theta$.  It is useful to
define the following relativistic correction factors that asymptote to
$1$ at large distance from the black hole:
\begin{equation}
\sA \equiv 1 + a^2/r^2 + 2 a^2/r^3,
\end{equation}
\begin{equation}
\sB \equiv 1 + a r^{-3/2},
\end{equation}
\begin{equation}
\sC \equiv 1 - 3/r + 2 a r^{-3/2},
\end{equation}
\begin{equation}
\sD \equiv 1 - 2/r + a^2/r^2,
\end{equation}
\begin{equation}
\sF \equiv 1 - 2 a r^{-3/2} + a^2/r^2,
\end{equation}
and
\begin{equation}
\sG \equiv 1 - 2/r + a r^{-3/2}.
\end{equation}
These definitions are identical to those used in \cite{nt73}.  

In what follows I will first develop a relativistic mechanical model for
the MRI (\S 2), then a local Cartesian coordinate system that permits
the treatment of the MRI near a rotating black hole using
nonrelativistic MHD (\S 3).  I compare the maximum growth of the
instability to the shear rate in \S 4, and give a brief summary and
prospects for future analysis in \S 5.

\section{A Mechanical Analogy}

BH considered the linear theory of a magnetized disk in the WKB
approximation.  The full theory is rather complicated, but \cite{bh92}
developed a useful mechanical model for the instability that consists of
two point masses on nearly circular orbit connected by a spring.  The
masses are analogous to fluid elements connected by a magnetic field,
which behaves like a spring.  This mechanical model captures key
features of the full MHD instability, including the maximum growth rate
and the fact that the instability grows by exchange of angular momentum.
Indeed, when both the magnetic field and the wavevector of the
perturbation are parallel to the axis of rotation the mechanical modes
have frequencies which are {\it identical} to the WKB modes of the full
MHD model if one replaces the natural frequency $\gamma$ of the
spring-mass system by the \alf frequency $|\bk \cdot \va|$.  It is
therefore natural to begin a study of the MRI in the Kerr metric by
considering a relativistic version of the \cite{bh92} mechanical model.

The orbits of free particles are described by the geodesic equation
\begin{equation}
{d^2 x^\mu\over{d \tau^2}} = - \Gamma^{\mu}_{\nu\lambda}
	{d x^\nu\over{d\tau}} {d x^\lambda\over{d\tau}}
\end{equation}
which has the following solution for circular equatorial orbits:
\begin{equation}
u^\mu = {d x^\nu\over{d\tau}} = 
	\{
	\sB \sC^{-1/2},
	0,
	0,
	r^{-3/2} \sC^{-1/2}
	\}
\end{equation}
\begin{equation}
= u^t \{1, 0, 0, \Omega\}
\end{equation}
where $\Omega = 1/(r^{3/2} + a) = r^{-3/2} \sB^{-1}$.

Consider perturbations about the circular orbit, $x^\mu \rightarrow
x^\mu + \xi^\mu$, where $\xi^\mu$ is small.  Then
\begin{equation}
{d^2 \xi^\mu\over{d \tau^2}} = -\partial_\alpha
\Gamma^{\mu}_{\nu\lambda}
	u^\nu u^\lambda \xi^\alpha 
	- 2 \Gamma^{\mu}_{\nu\lambda} u^\nu \xi^\lambda + \order(|{\bf
	  \xi}|^2).
\end{equation}
Substituting $\xi^\mu \sim \exp(-i\omega \tau)$ and evaluating the
connection coefficients in Boyer-Lindquist coordinates, one finds
\begin{equation}
\omega^4 (\omega^2 - \nu^2) (\omega^2 - \kappa^2) = 0.
\end{equation}
Here
\begin{equation}
\nu^2 = {1\over{r^3}} {1 - 4 a/r^{3/2} + 3 a^2/r^2 \over{\sC}}
\end{equation}
is the ``vertical'' frequency and
\begin{equation}
\kappa^2 = {1\over{r^3}} {1 - 6/r + 8 a r^{-3/2} - 3 a^2/r^2
	\over{\sC}}
\end{equation}
is the epicyclic frequency.  On the innermost stable circular orbit
(ISCO) $\kappa^2 = 0$.  The other four modes are zero-frequency
relabelings of the $t$ and $\phi$ coordinate.  

The frequencies $\nu$ and $\kappa$ are defined with respect to the
proper time $\tau$; they can be converted to frequencies with respect to
the Boyer-Lindquist time coordinate $t$ (and with respect to observers
at infinity) by dividing by $u^t = dt/d\tau = \sB \sC^{-1/2}$ for a
circular orbit.  Using subscript $t$ to denote a frequency with respect
to $t$,
\begin{equation}
\kappa_{t}^2 = \Omega^2 (1 - 6/r + 8 a r^{-3/2} - 3 a^2/r^2)
\end{equation}
\begin{equation}
\nu_{t}^2 = \Omega^2 (1 - 4 a r^{-3/2} + 3 a^2/r^2).
\end{equation}

Now consider two particles tied together by a spring.  The particles lie
at $x^\mu \pm \xi^\mu$, and in the absence of rotation the natural
frequency of the masses and spring is $\gamma$.  The particles orbits
now interact with the potential
\begin{equation}
{1\over{2}} \gamma^2 h_{\mu\nu} \xi^\mu \xi^\nu
\end{equation}
where $h_{\mu\nu} = g_{\mu\nu} + u_\mu u_\nu$ projects the displacements
into the space normal to the unperturbed (circular) orbit's four-velocity.  

The equations of motion can be derived in a variety of ways, perhaps
most simply by varying the action with respect to $\xi^\mu$.  The result
is
\begin{equation}
{d^2 \xi^\mu\over{d \tau^2}} = -\partial_\alpha
\Gamma^{\mu}_{\nu\lambda}
        u^\nu u^\lambda \xi^\alpha
        - 2 \Gamma^{\mu}_{\nu\lambda} u^\nu \xi^\lambda
	- \gamma^2 h^\mu_\nu \xi^\nu.
\end{equation}
Assuming that $\xi^\mu \sim \exp(-i\omega \tau)$, one finds that the
nontrivial frequencies that correspond to motions in the equatorial
plane obey
\begin{equation}\label{DISPA}
\omega^4 - \omega^2 (2 \gamma^2 + \kappa^2) + \gamma^2 (\gamma^2 - s^2)
= 0
\end{equation}
where
\begin{equation}
s^2 = {3\over{r^3}} {\sD\over{\sC}}.
\end{equation}
Equation (\ref{DISPA}) is quadratic in $\omega^2$ and so easily solved.
By varying $\gamma$ one finds that the maximum growth rate is
\begin{equation}
-\omega^2_{max} = {s^4\over{4 (s^2 + \kappa^2)}}
	= {9\over{16}} {1\over{r^3}} \left({\sD\over{\sC}}\right)^2
\end{equation}
which occurs for
\begin{equation}
\gamma^2 = {s^2\over{4}} {s^2 + 2 \kappa^2\over{s^2 + \kappa^2}}.
\end{equation}
The maximum growth rate is defined with respect to proper time $\tau$;
it is the growth rate measured by an observer on a circular orbit at
radius $r$.  Converting to frequencies with respect to Boyer-Lindquist
time $t$ (observers at infinity):
\begin{equation}
-\omega^2_{max,t} = {9\over{16}} \Omega^2 \left({\sD^2\over{\sC}}\right)
\end{equation}
This may be compared to the Newtonian maximum growth rate
$-\omega^2_{max,N} = (9/16) r^{-3}$.   I will refer to the ratio of the
relativistic maximum growth rate to Newtonian maximum growth rate $f
\equiv |\omega_{max}/\omega_{max,N}|$ as the {\it normalized growth
rate}.

Figure 1 shows the run of the normalized growth rate for $a = 0.9, 0,$
and $-0.9$ (i.e. a counterrotating disk).  The most dramatic effect is
the drop in normalized growth rate with respect to time $t$ in the
prograde disk.  This is due to gravitational redshift (the conversion
from time $\tau$ to time $t$) so it is a global feature of the spacetime
rather than a characteristic of the MRI: any disk instability would
experience the same redshift.  The effect is strongest on the ISCO for
prograde disks.  The normalized growth rate with respect to $t$ drops to
zero on the ISCO as $a \rightarrow 1$, since $u^t \rightarrow \infty$
there (see eq. 9).

The very limited variation of normalized growth rate with respect to
proper time $\tau$ as $r$ and $a$ vary {\it is} peculiar to the MRI.  
This can be understood as follows.  Since $-\omega^2_{max}
= s^4/(4 (s^2 + \kappa^2))$ and $\kappa^2 = 0$ at the ISCO,
$-\omega^2_{max} = s^2/4$ at the ISCO.  One may also show that $s^2 +
\kappa^2 = 4 r^{-3}$, hence $-\omega^2_{max} = r^{-3}$ at the ISCO.
Then $\omega^2_{max}/\omega^2_{max,N} = 16/9$ on the ISCO.  Since the
relativistic correction to the MRI growth rate is largest at the ISCO,
the normalized growth rate varies between $4/3$ at $r = r_{ISCO}$ and
$1$ as $r \rightarrow \infty$.

\section{Local Model}

In studies of thin disks in a Newtonian potential it has proven useful
to introduce a local Cartesian coordinate frame, or {\it local model}.
This frame is centered on a circular orbit, the $x$ axis points along
the radius vector, the $y$ axis points forward in azimuth, and the $z$
axis points up, normal to the disk.  Beginning with a global spherical
coordinate system $r,\theta,\phi$, the transformation to nonrelativistic
local coordinates is 
\begin{equation}
x = r - r_0
\end{equation}
\begin{equation}
y = r_0 (\phi - \phi_0 - \Omega t)
\end{equation}
where $\phi_0$ is a constant and here only $\Omega = r_0^{-3/2}$, and
\begin{equation}
z = r_0 ({\pi\over{2}} - \theta).
\end{equation}
Expanding to lowest order in $|{\bf x}|/r_0 \sim |\dot{\bf x}|/r_0\Omega$, 
the free particle equations of motion are
\begin{equation}
\ddot{x} = 2 \Omega \dot{y} + 3 \Omega^2 x,
\end{equation}
\begin{equation}
\ddot{y} = -2 \Omega \dot{x},
\end{equation}
\begin{equation}
\ddot{z} = -\Omega^2 z.
\end{equation}
The linear theory of magnetized disks in the WKB approximation can be
carried over in its entirety to this local model because higher order
terms in $\eps$ are also higher order terms in the WKB approximation.

In a relativistic disk in the equatorial plane of the Kerr metric we may
proceed in exactly the same way.  We want to erect local Cartesian
coordinates near a circular orbit.  In Boyer-Lindquist coordinates the
circular orbit is located at $r = r_0$ and has $d\phi/dt = \Omega_0 =
1/(r_0^{3/2} + a)$.  The transformation to local coordinates
$\tau,x,y,z$ is
\begin{equation}
\tau = t \sG \sC^{-1/2}
- (\phi - \phi_0) r_0^{1/2} \sF \sC^{-1/2}
\end{equation}
\begin{equation}
x = (r - r_0) \sD^{-1/2},
\end{equation}
\begin{equation}
y = r_0 (\phi - \phi_0 - \Omega_0 t) \sB \sD^{1/2} \sC^{-1/2}
\end{equation}
\begin{equation}
z = r_0 ({\pi\over{2}} - \theta).
\end{equation}
This transformation is determined by the conditions that the local
coordinates should be locally Minkowski, and that the point $x,y,z = 0$
should follow a circular orbit.  This transformation is (not by
accident) identical to the transformation to a corotating orthonormal
basis of one-forms given by \cite{nt73}.

Now consider the limit $\eps^2 \equiv (x^2 + y^2 + z^2)/r_0^2 \ll 1$.
Expanding $g_{tt}$ to $\order(\eps^2)$, $g_{ti}$ to $\order(\eps)$ and
$g_{ij}$ to $\order(1)$ (higher order terms are not
required for a consistent expansion of $ds^2$; one can think of $d\tau$
as being $\order(1)$ and $dx^i$ as being $\order(\eps)$), the metric
becomes
\begin{equation}
ds^2 = (-1 + s^2 x^2 - \nu^2 z^2) d\tau^2
	+ 4 \tO x d\tau dy + dx^2 + dy^2 + dz^2
\end{equation}
where $\tO = r_0^{-3/2}$,
\begin{equation}
s^2 = {3\over{r_0^3}} \left({\sD\over{\sC}}\right),
\end{equation}
and
\begin{equation}
\nu^2 = {1\over{r_0^3}} \left( {1 - 4 a r_0^{-3/2} + 3 a^2/r_0^2
\over{\sC}} \right).
\end{equation}
These are the same as the $s$ and $\nu$ defined in the last section.

The metric component $g_{\tau\tau}$ can be interpreted as $\approx -1 +
2\psi$ where $\psi$ is the effective gravitational potential, and the
$g_{\tau y}$ term indicates that the local model is a frame rotating
with frequency $\tO$.  Using this interpretation, or working directly
with the geodesic equation, one arrives at the equations of motion for a
free particle 
\begin{equation} \ddot{x} = 2 \tO \dot{y} + s^2 x,
\end{equation} 
\begin{equation} \ddot{y} = -2 \tO \dot{x},
\end{equation} 
\begin{equation} \ddot{z} = -\nu^2 z, 
\end{equation}
where we have assumed that $(\dot{x}^2 + \dot{y}^2 + \dot{z}^2)/(\tO
r_0^2) \sim \eps^2$.  One can look for motions with time dependence
$\exp(-i\omega\tau)$; the free particle frequencies are identical to
those found from the perturbed geodesic equation in \S 2.

If the disk is thin then the sound speed is small compared to the speed
of light.  Since one expects the fluid speeds within the disk to be
subsonic, the fluid motions are nonrelativistic.  It follows that in the
local Cartesian frame the fluid obeys the equations of nonrelativistic
fluid dynamics.  The MHD equations of motion in the local model are then
\begin{equation}
{D\bv\over{D t}} = -2 \tilde{\bf \Omega} \times \bv 
+ s^2 x {\bf e}_x - \nu^2 z {\bf e}_z
- {\bnabla p\over{\rho}} - {\bnabla B^2\over{8 \pi \rho}}
+ {({\bf B} \cdot \bnabla){\bf B}\over{4 \pi\rho}},
\end{equation}
where $\tilde{\bf \Omega} \equiv \tO {\bf e}_z$.  These equations,
coupled to the continuity equation, the (nonrelativistic) induction
equation, an energy equation, and the condition $\bnabla \cdot \bB =
0$, give a complete description of the local evolution of the plasma in
the MHD approximation.

The nonrelativistic linear analysis of the Balbus-Hawley instability can
now be carried over into the local frame by replacing the Newtonian
rotation frequency $\Omega$ with $\tO$, the coefficient of the tidal
expansion of the effective potential $(3/2) \Omega^2$ by $s^2$, and the
Newtonian vertical frequency $\Omega$ by $\nu$.  General relativistic
effects appear solely through the Coriolis parameter $\tO$ and the tidal
expansion parameter $s$.

To carry forward a linear analysis we need to specify an equilibrium.
The simplest choice, which captures most of the features of the
nonrelativistic instability, is a uniform vertical magnetic field near
the midplane of the disk, so the \alf velocity can be written $\bv_A =
(B/(4\pi\rho)^{1/2}) {\bf e}_z$.  I will restrict consideration to
perturbations of the form $\exp(-i\omega \tau + i k_z z)$ (more general
perturbations do not introduce qualitatively new features in ideal MHD).
Then the dispersion relation is

\begin{equation}
\omega^4 - \omega^2 (\kappa^2 + 2 (\bk \cdot \bv_A)^2) +
(\bk \cdot \bv_A)^2 ((\bk \cdot \bv_A)^2 - s^2) = 0
\end{equation}
Since this is identical to the relation obeyed by the mechanical model,
with $\gamma^2 \rightarrow (\bk \cdot \bv_A)^2$, the maximum growth rate
of the MHD instability is the same as that of the mechanical model.

\section{Relationship to Shear Rate}

\cite{bh92} conjectured that the maximum growth rate should always be
half the shear rate.  To check this, we need a relativistic
generalization of the shear rate.  The natural generalization involves
the scalar
\begin{equation}
\sigma^2 = \sigma^{\mu\nu} \sigma_{\mu\nu} 
\end{equation}
where
\begin{equation}
\sigma_{\alpha\beta} = {1\over{2}}\left(
        u_{\alpha;\mu} h^\mu_\beta + u_{\beta;\mu} h^\mu_\alpha
	        \right) - {1\over{3}}\Theta h_{\alpha\beta}
\end{equation}
is the rate-of-strain tensor and
\begin{equation}
\Theta \equiv {u^\alpha}_{;\alpha}
\end{equation}
is the divergence of the velocity field.  For a Cartesian shear flow
$v_y = A x$, where $A$ is constant, one finds $\sigma^2 = A^2/2$, so $2
\sigma^2$ is the square of the shear rate.

For circular equatorial geodesics in the Kerr metric $\sigma^2$ can be
evaluated in any frame, including the Boyer-Lindquist coordinate frame
(this is a lengthy calculation!).  \cite{nt73} evaluated
$\sigma_{\mu\nu}$ in an orthonormal tetrad moving with the flow, which
is equivalent to our local coordinate frame.  They found
\begin{equation}
\sigma_{(r)(\phi)} = \sigma_{(\phi)(r)} = 
{3\over{4}} {1\over{r^{3/2}}} \left( {\sD\over{\sC}} \right),
\end{equation}
and all other components are zero.  Then
\begin{equation}
\sigma^2 = {9\over{8}}{1\over{r^3}} \left({\sD\over{\sC}}\right).
\end{equation}
Evidently 
\begin{equation}
-\omega^2_{max} = {1\over{4}} (2 \sigma^2) 
\end{equation}
This confirms the relationship between the maximum growth rate and the
shear rate suggested by \cite{bh92}.  In the local coordinate system, of
course, the relationship is trivially verified.

\section{Conclusion}

I have shown that the growth rate of the MRI in thin, equatorial disks
around Kerr black holes does not differ sharply from its Newtonian
counterpart, at least as measured by an observer on a circular orbit.
As measured by an observer at infinity the growth rate is redshifted
strongly for instabilities growing near the marginally stable orbit in
corotating disks.

A similar analysis could be carried out for inclined disks.  The work of
\cite{bp75} and successors (see \citealt{di97} and references therein)
suggests, however, that close to the hole disks settle into the
equatorial plane of the black hole in the presence of angular momentum
diffusion (such as that which might result from MHD turbulence initiated
by the MRI).  At larger distance the deviations from Newtonian gravity
will be small and will likely not affect the development of the most
unstable modes, which grow on the dynamical timescale.

Earlier work by \cite{gp98} (see footnote 7) stated the results
presented here without proof.  \cite{ag02} gives an impressive fully
relativistic linear analysis that is equivalent to results presented
here (compare Figure 1 with his Figure 2).  The fully relativistic
analysis is perhaps the ultimate proof that the MRI carries over nearly
unchanged to relativistic disks; the present paper states these results
in a greatly simplified form.

\acknowledgments

This work was supported by NSF ITR grant PHY 02-05155 and by NSF PECASE
grant AST 00-93091.  I wish to thank John Hawley and an anonymous
referee for helpful comments.

\clearpage

\begin{figure}
\plottwo{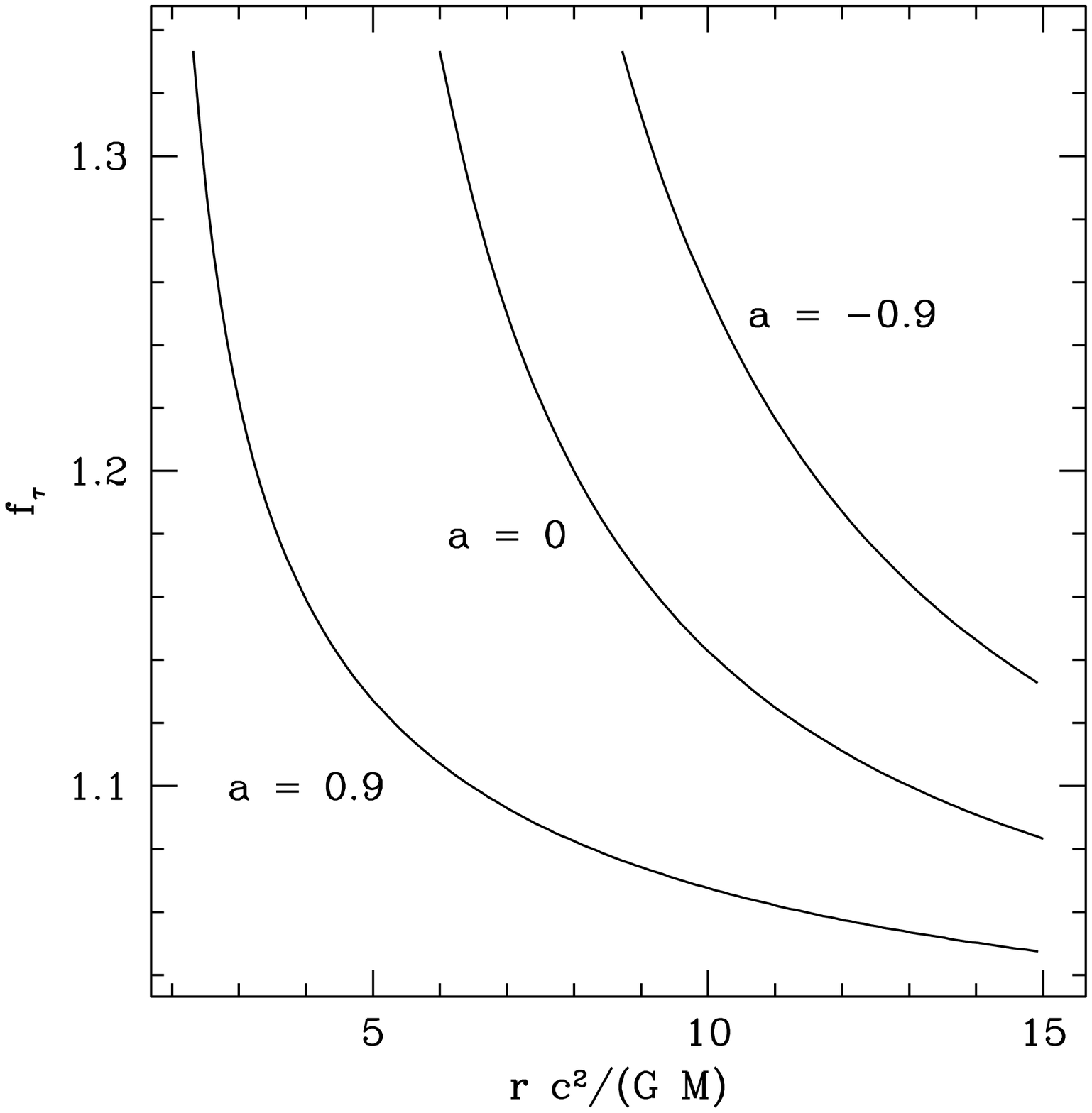}{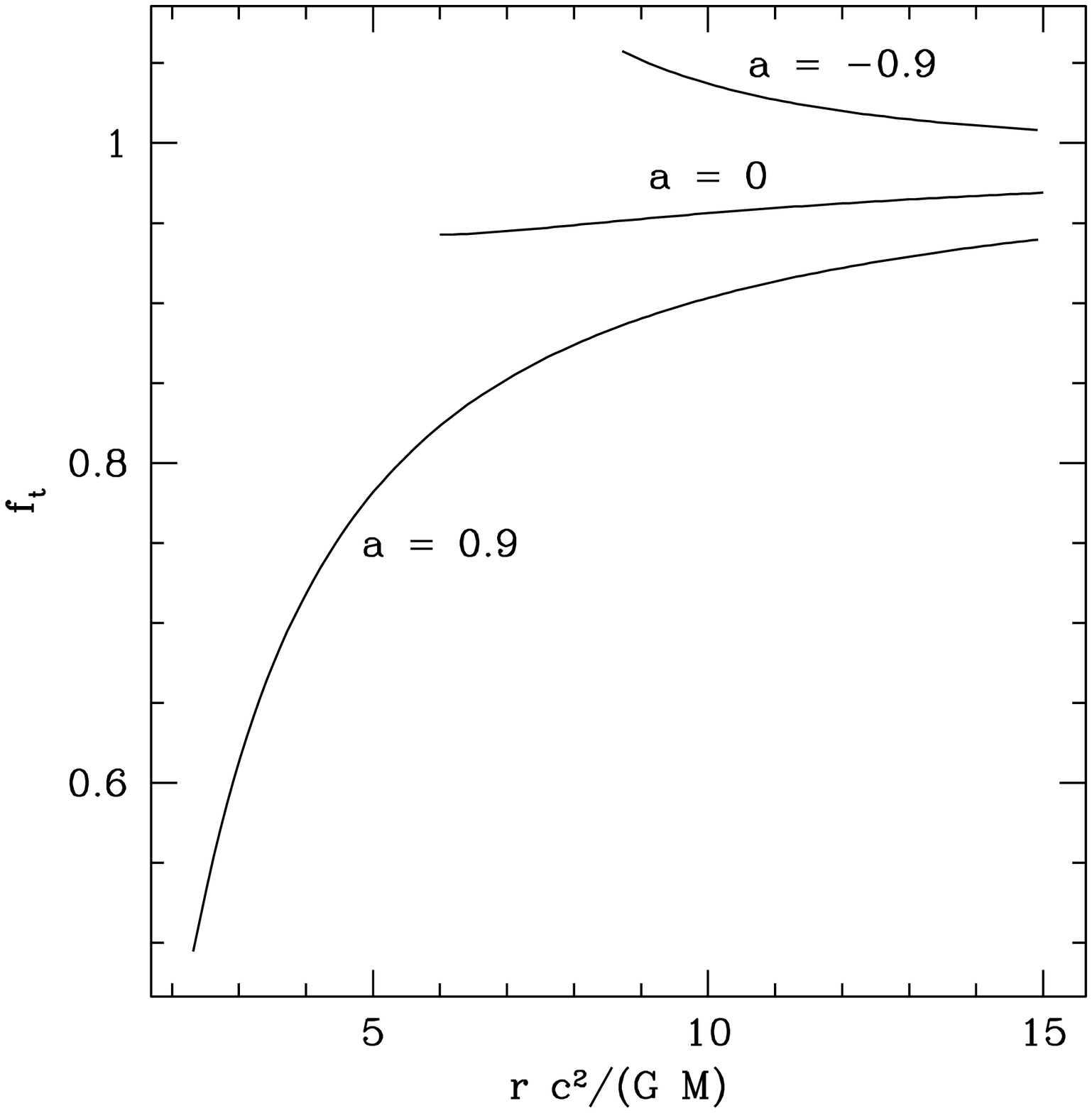}
\caption{
The figure shows the maximum growth rate of the Balbus-Hawley
instability for a thin disk in the equatorial plane of the Kerr metric
normalized to the naively calculated nonrelativistic growth rate.  The
left panel shows $f_\tau$, the normalized growth rate with respect to
the proper time $\tau$ of an observer on a circular orbit.  The right
panel shows $f_t$, the normalized growth rate with respect to the
Boyer-Lindquist time $t$, which is the growth rate measured by an
observer at large radius.
}
\end{figure}

\end{document}